\documentclass[12pt]{article}
\pdfoutput=1
\usepackage{amsmath,amssymb,amsfonts,amsthm,bm,bbm,cancel,wasysym}
\usepackage{epsfig,graphics,graphicx,epstopdf,caption,subcaption}
\graphicspath{{Charts/}}
\usepackage{array,booktabs,colortbl,colordvi,multirow}
\usepackage[numbers,sort&compress]{natbib}
\usepackage{colordvi,color,xcolor}
\usepackage{hyperref}
\usepackage{rotating}
\usepackage{comment}
\usepackage{feynmf}

\usepackage[symbol]{footmisc}
\renewcommand{\thefootnote}{\fnsymbol{footnote}}

\begin{document}

\begin{center}

{\Large {\bf Freeze-in Dark Matter via Lepton Portal: Hubble Tension and Stellar Cooling}}\\

\vspace*{0.75cm}

{Zixuan Xu$^1$, Shuai Xu$^2$, Ruopeng Zhang$^1$ and Sibo Zheng$^{1}$\footnote{Corresponding author: sibozheng.zju@gmail.com.}}\\

\vspace{0.5cm}
{$^1$Department of Physics, Chongqing University, Chongqing 401331, China\\
$^2$School of Physics and Telecommunications Engineering, Zhoukou Normal University, Henan 466001, China}
\end{center}
\vspace{.5cm}

\begin{abstract}
 \noindent 
 We propose a new freeze-in dark matter candidate which feebly couples to the standard model charged leptons. 
 The feeble interactions allow it (i) to freeze-in from the Standard Model thermal bath with its relic density being either a fraction or the entirety of the observed dark matter density and (ii) to radiatively decay to two photons in the dark matter mass ranges of order keV scale with lifetime larger than the age of Universe.
These features make this model a realistic realization of dark matter with late-time decay to reduce Hubble tension.
We show the best-fit value of $H_{0}=68.31(69.34)$ km~s$^{-1}$Mpc$^{-1}$ in light of Planck 2018+BAO(+LSS)+Pantheon data sets.
We then use stellar cooling data to place constraints on the parameter space favored by the Hubble tension. 
While the universal coupling scenario is excluded, the hierarchical coupling scenario can be tested by future observations of white dwarfs after a careful look into photon inverse decay, Primakoff and Bremsstrahlung emission of the dark matter in various stellar systems. 
The viable parameter space may be linked to anomalies in future X-ray telescopes.
\end{abstract}

\renewcommand{\thefootnote}{\arabic{footnote}}
\setcounter{footnote}{0}
\thispagestyle{empty}
\vfill
\newpage
\setcounter{page}{1}

\tableofcontents
\section{Introduction}
\label{sec1}
Searches of dark matter (DM) from the first direct detection made by \cite{Ahlen:1987mn} 
to cutting-edge experiments such as LZ \cite{LZ:2023poo}, XENONnT \cite{XENONCollaboration:2023orw},
DarkSide-50 \cite{DarkSide:2022knj} and SENSEI \cite{SENSEI:2020dpa} have significantly improved the exclusion limits on DM scattering cross sections either off nucleons or electrons for it being a weakly interacting massive particle (WIMP). 
The null experimental results on WIMP-like DM, which is motivated by a variety of new physics models beyond Standard Model (SM) related to electroweak symmetry breaking, initiate the studies of alternative DM scenarios. 

As a representative alternative, freeze-in DM can be produced in the early Universe through the freeze-in mechanism \cite{Hall:2009bx} as a result of feeble interactions with SM thermal bath, see \cite{Bernal:2017kxu} for a review.
So far, freeze-in DM via the standard model neutrino portal \cite{Asaka:2005cn,Becker:2018rve,Chianese:2018dsz,Datta:2021elq,Escudero:2019gvw, Arias-Aragon:2020qip, Huang:2021dba} and Higgs portal \cite{McDonald:2001vt, Kang:2015aqa} have been studied in the literature. 
In this study we propose a new freeze-in DM through the standard model charged lepton portal with the following Lagrangian
\begin{eqnarray}{\label{Lag}}
 \mathcal{L}_{\phi}=\frac{1}{2}(\partial\phi)^{2}-\frac{1}{2}m^{2}_{\phi}\phi^{2}-\lambda_{\ell}\bar{\ell}\ell\phi,
\end{eqnarray}
where $\ell=\{e,\nu,\tau\}$ are the charged leptons, $\phi$ is a dark scalar degree of freedom with mass $m_{\phi}$ without either baryon or lepton number, and $\lambda_{\ell}$ is the coupling constant.\footnote{Although explicit realizations of this effective interaction are beyond the scope of this study, it can be constructed e.g., via coupling $\phi$ to vector-like fermions \cite{Kumar:2015tna} that mix with the SM leptons, where the feeble coupling reads as $\lambda_{\ell}\sim \epsilon^{2}(\upsilon/M_{\ell})^{2}$ with $\epsilon$, $\upsilon$ and $M_\ell$ referring to the small mixing angle, the electroweak scale and the vectorlike lepton mass respectively. As shown in \cite{Kumar:2015tna}, $M_\ell$ below $\sim 200$ GeV has been excluded by the LHC searches on multi-lepton final states. \\}  
Note, a feeble coupling $\lambda_{\ell}$ makes this model different from the WIMP-like DM studied by \cite{Feldstein:2013kka, Bai:2014osa} from a viewpoint of phenomenology. 

Along with the freeze-in production, another key feature of this model is the late-time decay of $\phi$,
either through a tree-level channel $\phi\rightarrow\ell\bar{\ell}$ in the mass region $m_{\phi}>2m_{\ell}$ or a loop-level channel $\phi\rightarrow\gamma\gamma$ in the mass region $m_{\phi}<2m_{\ell}$.
DM with a late-time decay is of interest in cosmic ray anomalies \cite{Feldstein:2013kka, Nardi:2008ix, Cirelli:2009dv}, 
an excess of recoil electrons at XENON1T \cite{Choi:2020udy,Xu:2020qsy, Dutta:2021nsy}, 
and Hubble tension \cite{Enqvist:2015ara,Poulin:2016nat, Audren:2014bca, Anchordoqui:2015lqa,Bringmann:2018jpr,Holm:2022eqq, Kumar:2018yhh,Pandey:2019plg,Vattis:2019efj,Xiao:2019ccl,Clark:2020miy,Nygaard:2020sow,FrancoAbellan:2021sxk,Simon:2022ftd,Alvi:2022aam}.
The Hubble tension is a $\sim 5\sigma$ discrepancy between the measurements on Hubble parameter $H_0$ derived from Planck \cite{Aghanim:2018eyx} and local experiments \cite{Riess:2020fzl,Riess:2021jrx}.
For recent reviews on this topic see \cite{DiValentino:2021izs,Schoneberg:2021qvd}.
Unlike in \cite{Enqvist:2015ara,Poulin:2016nat, Audren:2014bca, Anchordoqui:2015lqa,Bringmann:2018jpr,Holm:2022eqq, Kumar:2018yhh,Pandey:2019plg,Vattis:2019efj,Xiao:2019ccl,Clark:2020miy,Nygaard:2020sow,FrancoAbellan:2021sxk,Simon:2022ftd,Alvi:2022aam}, our model serves as a more concrete example of freeze-in DM with late-time decay, 
as $\phi$ decay width $\Gamma_{\phi}$ and energy density $\rho_{\phi}$ are no longer two independent variables but instead correlated in terms of the model parameters $m_{\phi}$ and $\lambda_{\ell}$ in Eq.(\ref{Lag}).
Therefore, we can test the explicit parameter space which reduces the Hubble tension by complimentary cosmological or astrophysical experiments. 

The aims of this study are two-fold. 
The first task is to make Markov Chain Monte Carlo (MCMC) analysis of our model in light of cosmological data sets.
While the cosmological constraints arising from Cosmic Microwave Background (CMB) and Large-scale Structure (LSS) can be accommodated in the course of MCMC analysis as in the aforementioned studies,
how to detect the freeze-in DM model is rather challenging. 
In practice, due to the feeble couplings and the light mass $\phi$ should have a role to play in various astrophysical stellar systems \cite{Raffelt1996} similar to light axion, dark photon or mini-charged particles. 
The second task of this work is to use stellar cooling data, some of which can be very precise, to place constraints on the feeble couplings. 
These constraints are expected to be more stringent than current ground-based experiments. 

The rest of the paper is organized as follows. 
In Sec.\ref{M} we calculate the relic abundance of $\phi$ via the freeze-in mechanism,
which will be specifically divided into two cases, i.e, the light mass range $m_{\phi}<2m_{\ell}$ and the heavy mass range $m_{\phi}>2m_{\ell}$ corresponding to the loop- and tree-level inverse decay respectively. 
We will show that the relic density can fully or partially accommodate the observed DM density while the lifetime is larger than the age of Universe in the case of light mass range.
In Sec.\ref{Cos} we firstly discuss the impacts of the late-time decays of DM into photons on cosmological observables such as the  CMB and matter power spectra, then address the parameter space which reduces the Hubble tension in terms of MCMC fit to two different cosmological data sets.
Afterward, we explicitly explore the photon inverse decay, Primakoff and Bremsstrahlung emission of $\phi$ in various stellar systems.
It tuns out that current stellar cooling data has excluded the parameter space within the universal coupling scenario while that of the hierarchical coupling scenario can be tested by future observations of white dwarfs.
Finally we conclude in Sec.\ref{con}.

\section{The dark matter model}
\label{M}
In this section we discuss the relic abundance of scalar $\phi$ in the model defined as in Eq.(\ref{Lag}), 
which is subject to freeze-in and subsequent decay.

The freeze-in production of $\phi$ particle is through the inverse decay process $\ell^{-}\ell^{+}\rightarrow \phi$.
This occurs whenever $m_{\phi}$ is bigger than center-of-mass energy of the two incoming leptons, i.e, 
$m_{\phi}\geq 2m_{\ell}$. 
On the contrary, in the mass ranges of $m_{\phi}< 2m_{\ell}$ the tree-level inverse decay is replaced by one-loop analogy $\gamma\gamma\rightarrow\phi$ via $\ell$ triangle diagram.\footnote{It corresponds to an effective operator $(\phi/m_{\ell})F^{2}$ with $F$ being the electromagnetic field strength.
Likewise, a radiative decay $\phi\rightarrow \nu_{\ell}\bar{\nu}_{\ell}$ can be produced as well. 
Since the width of $\phi\rightarrow \nu_{\ell}\bar{\nu}_{\ell}$ is much smaller than that of $\phi\rightarrow \gamma\gamma$, 
we neglect the inverse decay due to neutrinos.}
Either of these two inverse decays contributes to the number density of $\phi$ particle as \cite{Hall:2009bx}  
\begin{eqnarray}{\label{dc}}
\dot{n}_{\phi}+3Hn_{\phi}=\int d\Pi_{\phi}d\Pi_{X}d\Pi_{Y} (2\pi)^{4} \delta^{(4)}(p_{\phi}-p_{X}-p_{Y})\mid M\mid_{X+Y\rightarrow \phi}^{2} f_{X}f_{Y}
 \end{eqnarray}
where $X$, $Y$ refer to $\ell$ ($\gamma$) in the tree (loop)-level decay,
$\Pi$s are phase space elements \cite{Edsjo:1997bg} and $f$s are phase space densities.
Here, $\mid M\mid^{2}_{X+Y\rightarrow \phi}$ is squared amplitude of the inverse decay $X+Y\rightarrow \phi$,
which is equal to $\mid M\mid^{2}_{\phi\rightarrow X+Y}$ in the case of CP conservation as we assume.
Solving Eq.(\ref{dc}) in terms of a new variable $Y_{\phi}\equiv n_{\phi}/s$ with $s$ entropy density of thermal bath gives the relic density  \cite{Hall:2009bx} 
\begin{eqnarray}{\label{drelic}}
\Omega_{\phi}h^{2}\mid_{2\rightarrow1}\sim \frac{10^{27}}{g_{s}\sqrt{g_{\rho}}}\frac{\Gamma_{\phi}}{m_{\phi}},
 \end{eqnarray}
 where $g_{s}$ and $g_{\rho}$ are the number of degrees of freedom in entropy and energy density respectively, and
\begin{equation}{\label{dw}}
\Gamma_{\phi}\approx
 \left\{
\begin{array}{lcl}
\frac{\lambda^{2}_{\ell}}{8\pi}m_{\phi}, ~~~~~~~~~~~~~~~~~~~~~~ m_{\phi}> 2m_{\ell},\\
\frac{\alpha^{2}\lambda^{2}_{\ell}}{256\pi^{3}}\frac{m^{3}_{\phi}}{m^{2}_{\ell}}\mid F\left(\frac{4m^{2}_{\ell}}{m^{2}_{\phi}}\right)\mid^{2}, ~~ m_{\phi}<2m_{\ell},\\
\end{array}\right.
\end{equation}
 is the decay width with $F(x)=-2x\{1+(1-x)[\sin^{-1}(\sqrt{1/x})]^{2}\}$ \cite{Ellis:1975ap,Shifman:1979eb} approximately equal to $-4/3$ in the large $x$ limit. 
There is a sum over flavor in Eq.(\ref{dw}) if needed. 

In addition, the freeze-in production of $\phi$ particles can be through annihilation processes such as $\ell^{+}\ell^{-}\rightarrow\gamma\phi$.
Such processes contribute to the number density as follows \cite{Hall:2009bx}  
\begin{eqnarray}{\label{ac}}
\dot{n}_{\phi}+3Hn_{\phi}=\int d\Pi_{\phi}d\Pi_{X}d\Pi_{Y}\Pi_{Z}(2\pi)^{4} \delta^{(4)}(p_{Z}+p_{\phi}-p_{X}-p_{Y})\mid M\mid_{X+Y\rightarrow Z\phi}^{2} f_{X}f_{Y},
 \end{eqnarray}
where $X$, $Y$ and $Z$ denote $\ell$ and $\gamma$. 
Under the limit $m_{\phi}<<m_{\ell}$ the $2\rightarrow 2$ contribution in Eq.(\ref{ac}) is given by
\begin{eqnarray}{\label{drelic2}}
\Omega_{\phi}h^{2}\mid_{2\rightarrow2}\sim \frac{10^{27}}{g_{s}\sqrt{g_{\rho}}} \alpha\lambda^{2}_{\ell}\frac{m_{\phi}}{m_{\ell}},
 \end{eqnarray}
which is the same order of $\lambda^{2}_{\ell}$ as the $2\rightarrow 1$ process in Eq.(\ref{drelic}). 
Compared to the $2\rightarrow 1$ contribution, the $2\rightarrow 2$ contribution is subdominant (dominant) in the case with $m_{\phi}>2m_{\ell}$ ($m_{\phi}<2m_{\ell}$),
verified by a numerical analysis via the publicly available code micrOMEGAs5.0 \cite{Belanger:2018ccd} adopted in this work.

In terms of the freeze-in abundance, one obtains the present relic density of $\phi$
\begin{eqnarray}{\label{frelic}}
\Omega_{\phi}h^{2}=\Omega_{\phi}h^{2}\mid_{t_{*}}\exp[-\Gamma_{\phi}(t_{0}-t_{*})],
\end{eqnarray}
where $\Omega_{\phi}h^{2}\mid_{t_{*}}$ is the freeze-in relic abundance with $t_*$ the end time of freeze-in process and $t_{0}\approx 13.78$ Gyr is the age of Universe.
If $\phi$'s lifetime $\tau_{\phi}=\Gamma^{-1}_{\phi}$ is at least a few times larger than $t_{0}$ the $e$-factor in Eq.(\ref{frelic}) can be simply neglected.
In contrast, if $\tau_{\phi}$ is far smaller than $t_{0}$,
the $e$-factor takes over the freeze-in contribution and makes $\phi$ be irrelevant in the evolution of Universe. 
To serve our purpose we consider $\phi$ satisfying the following two conditions:
\begin{itemize}
\item the fraction parameter $f_{\phi}\equiv\Omega_{\phi}/\Omega^{\rm{ini}}_{\rm{cdm}}$ is less than unity, 
with $\Omega^{\rm{ini}}_{\rm{cdm}}$ the value of DM relic density reported by Planck 2018 data \cite{Aghanim:2018eyx}.
\item the lifetime $\tau_{\phi}$ is larger than the age of Universe $t_0$.
\end{itemize}
With these two features, $\phi$ is a natural realization of DM with late-time decay.

\subsection{$m_{\phi}<2m_{\ell}$}

\begin{figure}
\centering
\includegraphics[width=8cm,height=8cm]{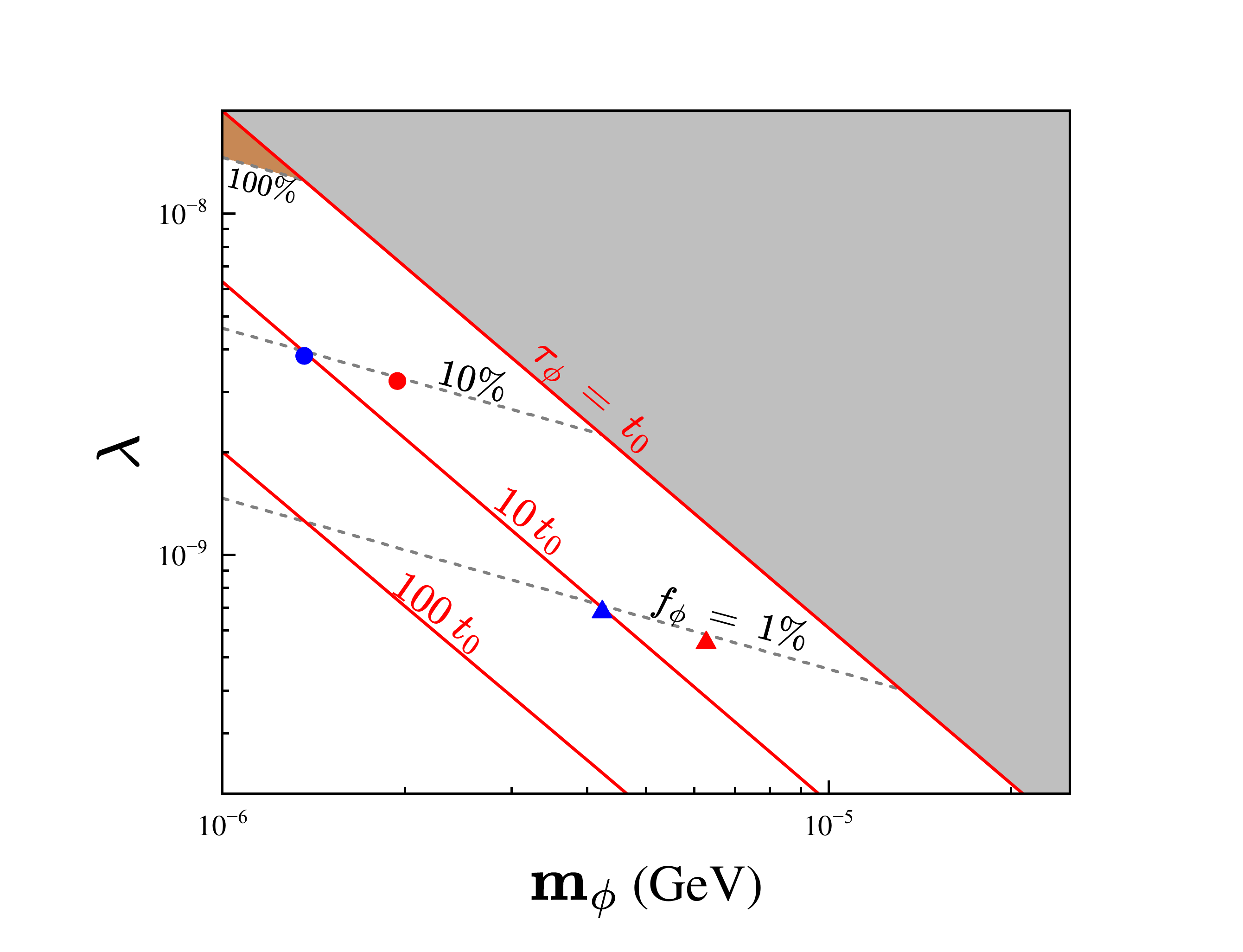}
\includegraphics[width=8cm,height=8cm]{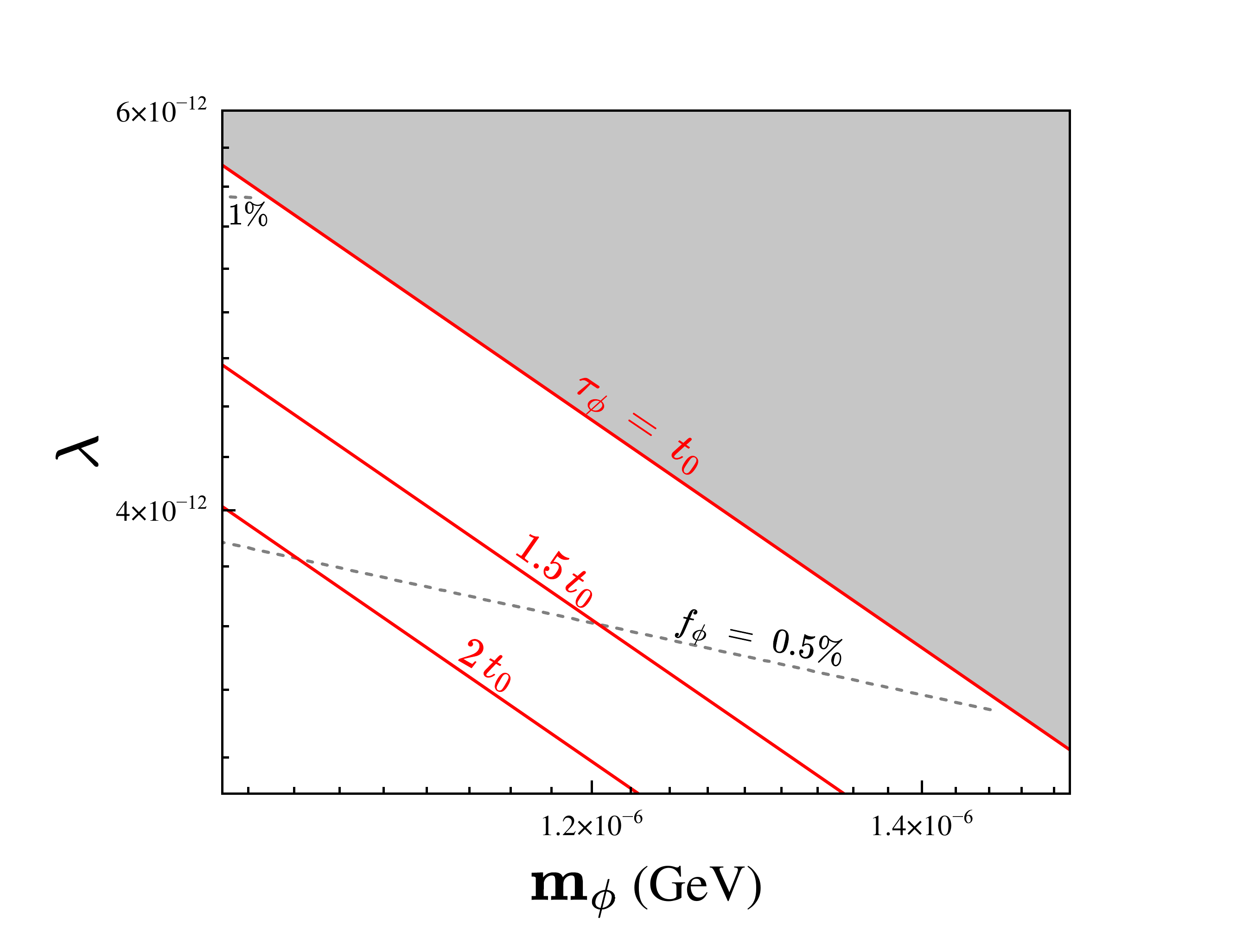}
\centering
\caption{Relic abundance of $\phi$ in the mass region with $1~\rm{keV}<m_{\phi}<2m_{\ell}$ in the $\mathbf{hierarchical}$ coupling scenario with $\lambda_{e}<<\lambda_{\mu}<<\lambda_{\tau}=\lambda$ (left) and the $\mathbf{universal}$ coupling scenario with $\lambda_{e}\approx\lambda_{\mu}\approx\lambda_{\tau}=\lambda$ (right).
Contours of $f_\phi$ and  $\tau_{\phi}$ (in units of $t_0$) are shown in dashed and solid respectively.  
Shaded gray region is excluded by overproduction, while shaded purple region is excluded by $\tau_{\phi}<t_{0}$.
In the left panel the four benchmark points will be used in Fig.\ref{Cl} and Fig.\ref{Pk}.}
\label{rlight}
\end{figure}

\begin{figure}
\centering
\includegraphics[width=8cm,height=8cm]{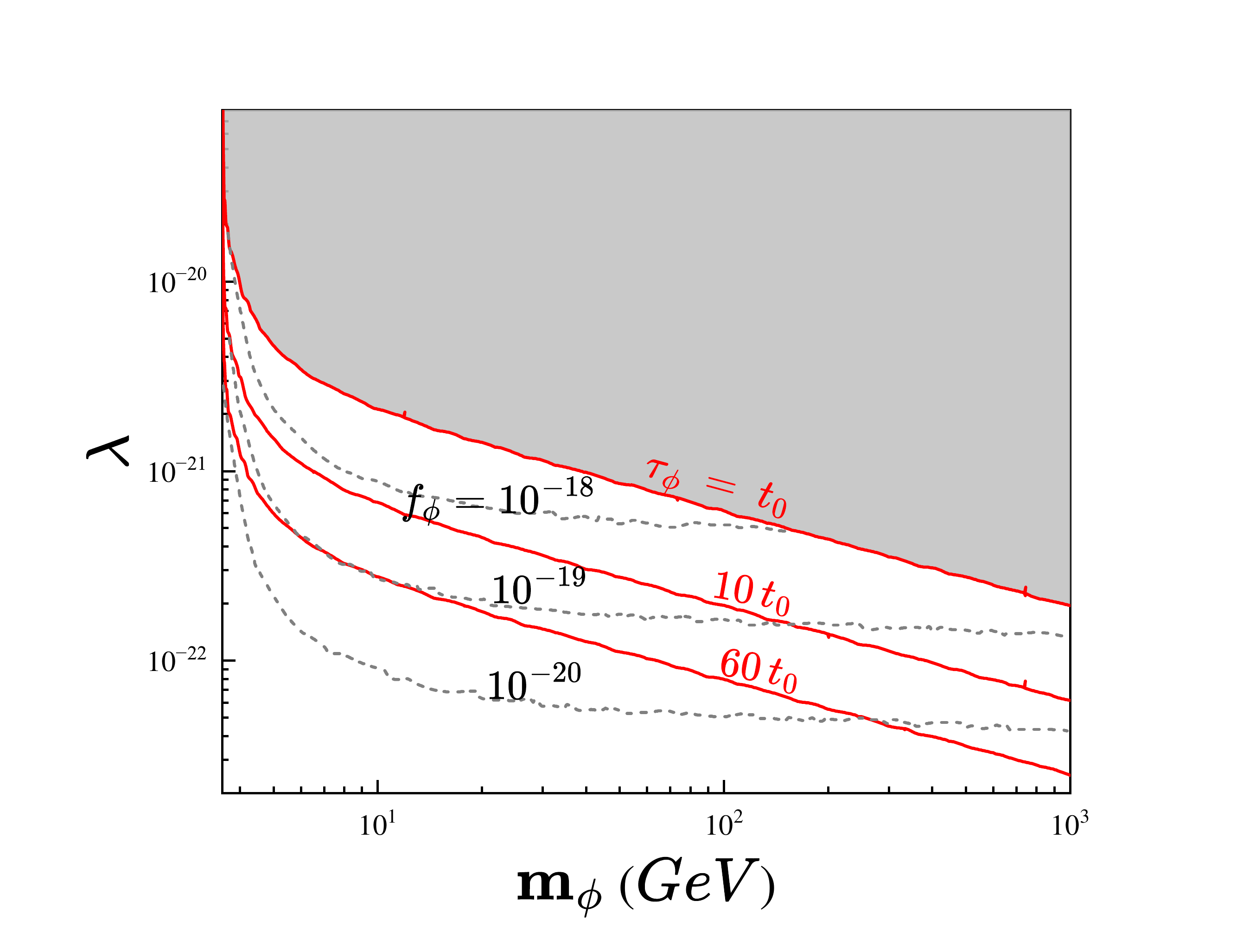}
\includegraphics[width=8cm,height=8cm]{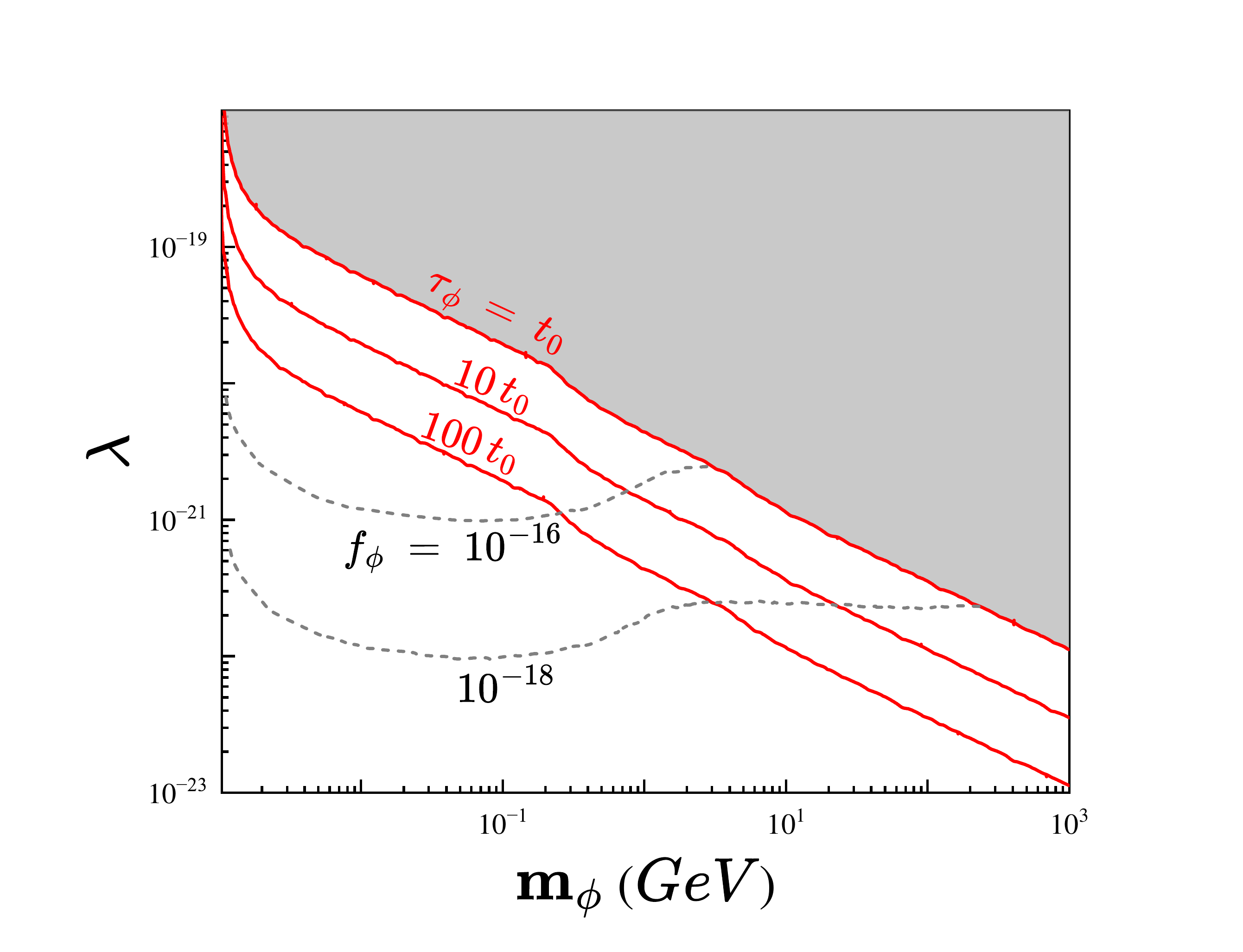}
\centering
\caption{Relic abundance of $\phi$ in the mass region with $m_{\phi}>2m_{\ell}$ in the $\mathbf{hierarchical}$ (left) and $\mathbf{universal}$ (right) coupling scenario respectively. Contours of $f_\phi$ and $\tau_{\phi}$ in units of $t_0$ are shown in dashed and solid respectively. Shaded gray region is excluded by $\tau_{\phi}<t_{0}$.}
\label{rHeavy}
\end{figure}

$\mathbf{Hierarchical}$ coupling scenario. 
We firstly consider the case of the hierarchical coupling scenario with $\lambda_{e}<<\lambda_{\mu}<<\lambda_{\tau}=\lambda$.
The left panel of Fig.\ref{rlight} shows the relic abundance of $\phi$ projected to the plane of $m_{\phi}-\lambda$ in the mass region $1~\rm{keV}<m_{\phi}<2m_{\tau}$. 
Explicitly, we show the contours of $f_{\phi}$ and $\tau_{\phi}$ in units of $t_0$ in dashed and solid respectively. 
The highlighted regions point to $f_{\phi}\sim 1\%-100\%$ and $\tau_{\phi}/t_{0}\sim 1-10^{2}$.
In this figure the shaded gray (purple) region is excluded by overproduction ($\tau_{\phi}<t_{0}$).

$\mathbf{Universal}$ coupling scenario. We now consider the case of the universal coupling scenario with $\lambda_{e}\approx\lambda_{\mu}\approx\lambda_{\tau}=\lambda$.
Compared to the hierarchical coupling scenario, 
each previous subprocess is now replaced by three copies of it, i.e, $\Omega h^{2}\mid_{2\rightarrow 1}\rightarrow \sum_{i}\left(\Omega h^{2}\mid_{2\rightarrow 1}\right)_{i}$, with $i=\{e,\mu,\tau\}$. 
Meanwhile, the mass range is now $m_{\phi}<2m_{e}$ instead of $m_{\phi}<2m_{\tau}$.
The right panel of Fig.\ref{rlight} presents the relic abundance of $\phi$ projected to the plane of $m_{\phi}-\lambda$ in the mass region $1~\rm{keV}<m_{\phi}<2m_{e}$,
where the contours of $f_{\phi}$ and $\tau_{\phi}$ are illustrated in the same way as in the left panel.
Similar to the hierarchical coupling scenario, 
the highlighted regions point to $f_{\phi}\sim 0.1\%-1\%$ and $\tau_{\phi}/t_{0}\sim 1-10$.

Here, $m_{\phi}$ less than $1$ keV has not been considered, which has been excluded by the Lyman-$\alpha $ constraint \cite{Garzilli:2019qki,Villasenor:2022aiy,Capozzi:2023xie} for $f_{\phi}=100\%$.

\subsection{$m_{\phi}>2m_{\ell}$}
\label{treedecay}
As seen in Eq.(\ref{dw}), $\Gamma_\phi$ in the mass range with $m_{\phi}>2m_{\ell}$ is dominated by the tree-level decay instead of radiative decay as studied above.
The requirement $\tau_{\phi}\geq t_{0}$ immediately implies that the magnitude of $\lambda$ is smaller than $\sim 10^{-20}-10^{-19}$.
This is explicitly shown in Fig.\ref{rHeavy} where we plot contours of $\phi$ relic abundance and $\tau_{\phi}$,
with the left and right panel therein corresponding to the $\mathbf{hierarchical}$ and $\mathbf{universal}$ coupling scenario respectively.
Either in the hierarchical or universal coupling scenario 
the small $\lambda$ now results in a negligible relic abundance in the parameter regions with $\tau_{\phi}\geq t_{0}$, 
compared to the case of radiative decay as shown in Fig.\ref{rlight}.
We would like to remind the reader that  in the universal coupling scenario
$\phi$ decays to both $\gamma$ and $\ell$ in the mass range of $m_{e}<m_{\phi}<m_{\tau}$ corresponding to the radiative and tree-level decay respectively, with $\ell$ at least composed of $e$.

\section{Cosmological constraints}
\label{Cos}
The late-time decay of DM affects both  \emph{background}  and \emph{perturbations} of cosmological surveys. 
In this section we focus on the DM model with $m_{\phi}<2m_{\ell}$ and the hierarchical coupling scenario for illustration. 
The reason for neglecting the universal coupling scenario will be explained in Sec.\ref{astro}.

The impacts on the \emph{background}  directly follow the background equations of DM energy density $\rho_{\phi}$ and radiation energy density $\rho_{r}$ \cite{Poulin:2016nat,Kumar:2018yhh}
\begin{eqnarray}{\label{classical}}
\rho'_{\phi}+3\frac{a'}{a}\rho_{\phi}&=&-a\Gamma_{\phi}\rho_{\phi},  \nonumber\\
\rho'_{r}+4\frac{a'}{a}\rho_{r}&=&a\Gamma_{\phi}\rho_{\phi}.
 \end{eqnarray}
 where primes denote derivatives with respect to conformal time and $a$ is the scale factor.

The effects of the late-time decay on the \emph{perturbations} can be derived from the linear perturbation equations of the radiation and DM denoted by $\delta_{r}$ and $\delta_{\phi}$ respectively in synchronous gauge as 
\begin{eqnarray}{\label{quantum}}
\delta'_{\phi}+\frac{h'}{2}&=&0,\nonumber\\
\delta'_{r}+\frac{4}{3}\theta_{r}+\frac{2}{3}h'&=&a\Gamma_{\phi}\frac{\rho_{\phi}}{\rho_{r}}(\delta_{\phi}-\delta_{r}),  \nonumber\\
\theta'_{r}-\frac{k^{2}}{4}(\delta_{r}-4\sigma_{r})-an_{e}\sigma_{T}(\theta_{b}-\theta_{r})&=&-a\Gamma_{\phi}\frac{\rho_{\phi}}{\rho_{r}}\theta_{r},  \\
\sigma'_{r}-\frac{4}{15}\theta_{r}-\frac{2}{15}h'-\frac{4}{5}\eta'+\frac{3}{10}kF_{3}+\frac{9}{10}an_{e}\sigma_{T}\sigma_{r}-\frac{1}{20}an_{e}\sigma_{T}(G_{0}+G_{2})&=&-a\Gamma_{\phi}\frac{\rho_{\phi}}{\rho_{r}}\sigma_{r},\nonumber\\
F'_{\ell}+\frac{k}{2\ell+1}[(\ell+1)F_{\ell+1}-\ell F_{\ell-1}]&=&-an_{e}\sigma_{T}F_{\ell},\  \ell \geq 3 \nonumber
\end{eqnarray}
where the results at the order $\ell =1$ \cite{Poulin:2016nat,Kumar:2018yhh} have been extended to $\ell\geq 2$ according to \cite{Kaplinghat:1999,Ma:1995},
with $k$ the wavenumber, $h$ one of the two scalar modes in this gauge,
$\theta_{b}$ the divergence of baryon fluid, 
$\sigma_{T}$ the Thomson scattering cross section referring to collision between the photon and baryon fluid before recombination,
and $F_{\ell}$ and $G_{\ell}$ defined in \cite{Ma:1995,Audren:2014bca,Poulin:2016nat}.
Without the decay terms in Eq.(\ref{quantum}) one returns to the well-known standard continuity and Euler equations of photon fluid. 

Note that $\Gamma_\phi$ in Eqs.(\ref{classical})-(\ref{quantum})  has been assumed to be dominated by the decay of $\phi\rightarrow \gamma\gamma$ such as in the situation with $m_{\phi}<2m_{\ell}$ and the $\mathbf{hierarchical}$ couplings. If not so, a branching ratio should be properly taken into account.

\subsection{Effects on cosmological observables}
We focus on the effects of the late-time DM decay on the CMB power spectrum $C^{XY}_{\ell}$ with $X, Y=\{T, E\}$ and the DM power spectrum $P(k)$.
This topic has been studied in various contexts such as 
DM decaying to photons/dark radiation \cite{Audren:2014bca,Anchordoqui:2015lqa,Poulin:2016nat,Bringmann:2018jpr,Holm:2022eqq, Kumar:2018yhh,Pandey:2019plg,Vattis:2019efj,Xiao:2019ccl,Clark:2020miy,Nygaard:2020sow,FrancoAbellan:2021sxk,Simon:2022ftd,Alvi:2022aam} under certain assumptions.
Especially, in these studies the decaying DM energy density and the decay width have to be considered as two independent input parameters for CLASS \cite{Lesgourgues:2011re,Blas:2011rf} to solve Eq.(\ref{quantum}).
As emphasized above,
they are actually correlated to each other in an explicit model e.g., in terms of the fundamental model parameters  $m_\phi$ and $\lambda$ in our case.
Therefore, we will directly use the model parameters as the inputs by embedding micrOMEGAs5.0 into CLASS.

In this subsection we adopt the best-fit values of Planck 2018 data \cite{Aghanim:2018eyx} on the cosmological parameters: 
$\Omega_{b}h^{2}=0.022383$, $\Omega^{\rm{ini}}_{\rm{cdm}}h^{2}=\left(\Omega_{\rm{sdm}}+\Omega_{\rm{\phi}}\mid_{t_{*}}\right)h^{2}=0.12011$, $100\theta_{s}=1.041085$, 
$n_{s}=0.9660$, $\ln\left(10^{10}A_{s}\right)=3.0448$ and $\tau_{\rm{reio}}=0.0543$, 
where $\Omega_{\rm{sdm}}$ is the energy density of stable DM component.

\begin{figure}
\centering
\includegraphics[width=8cm,height=8cm]{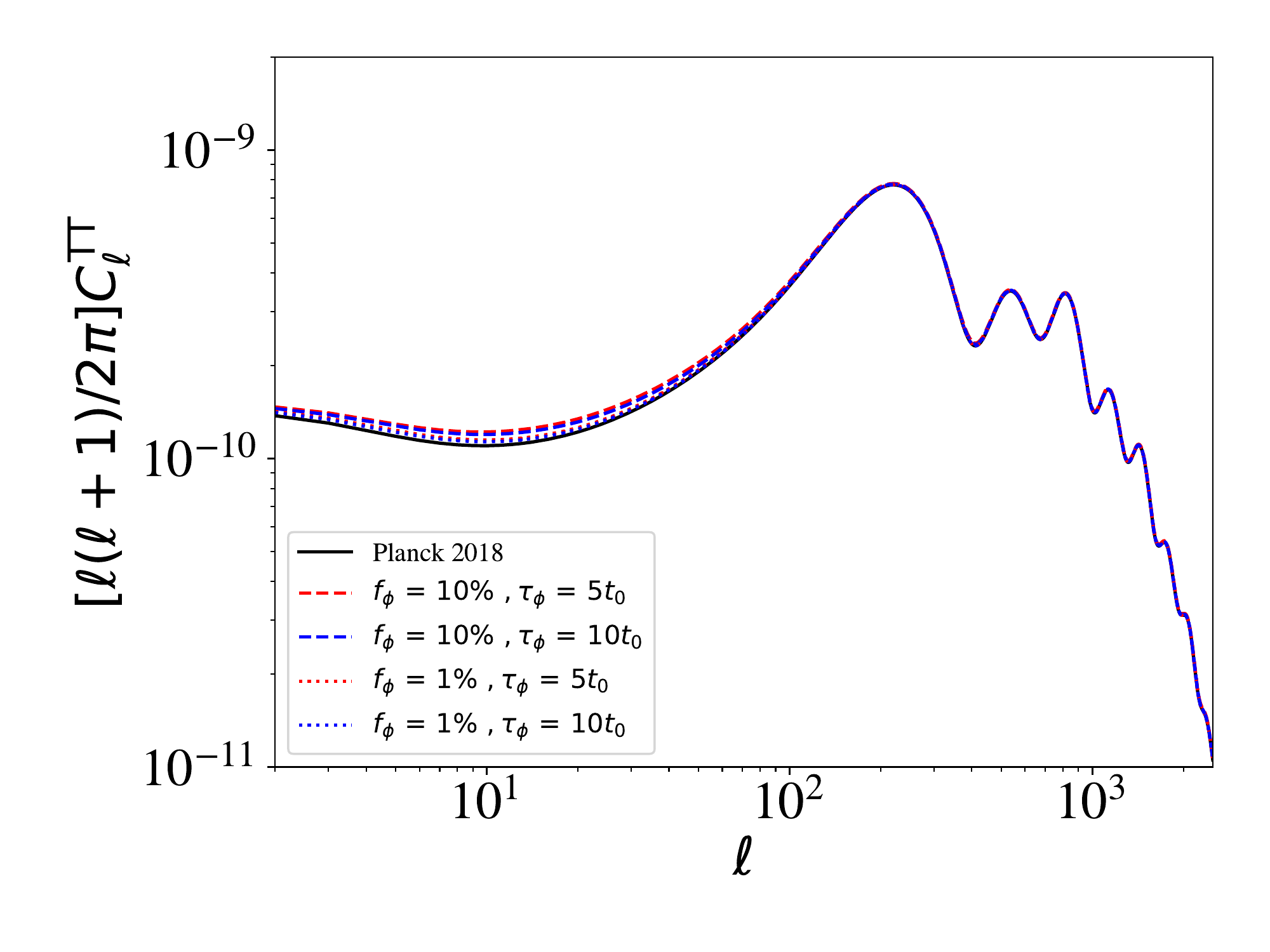}
\includegraphics[width=8cm,height=8cm]{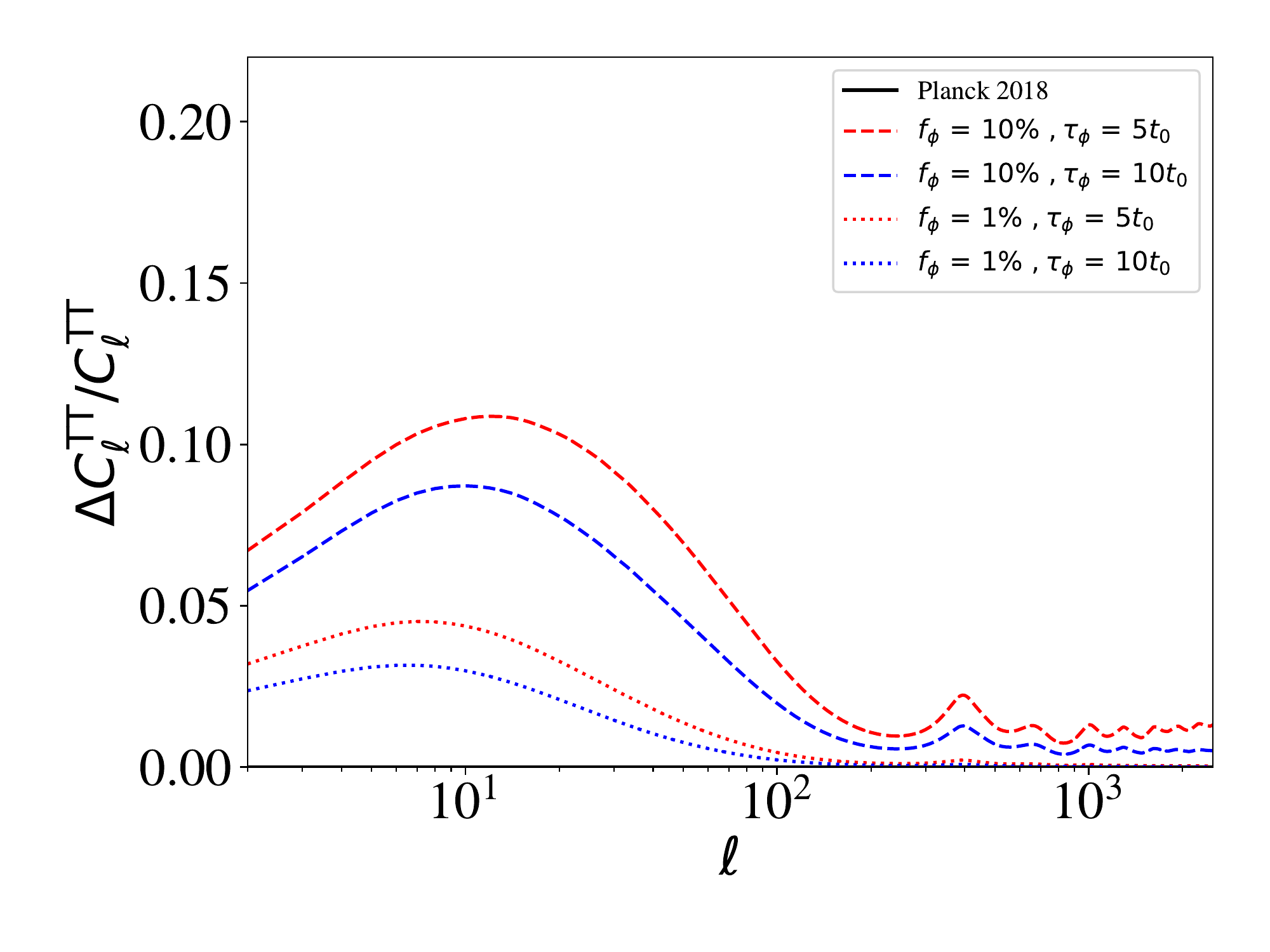}
\centering
\caption{Left: CMB temperature power spectrum compared to Planck 2018 $\Lambda$CDM (black) for the four benchmark points as shown in the right panel. Right: magnitudes of the deviation over the $\Lambda$CDM value.}
\label{Cl}
\end{figure}

\begin{figure}
\centering
\includegraphics[width=8cm,height=8cm]{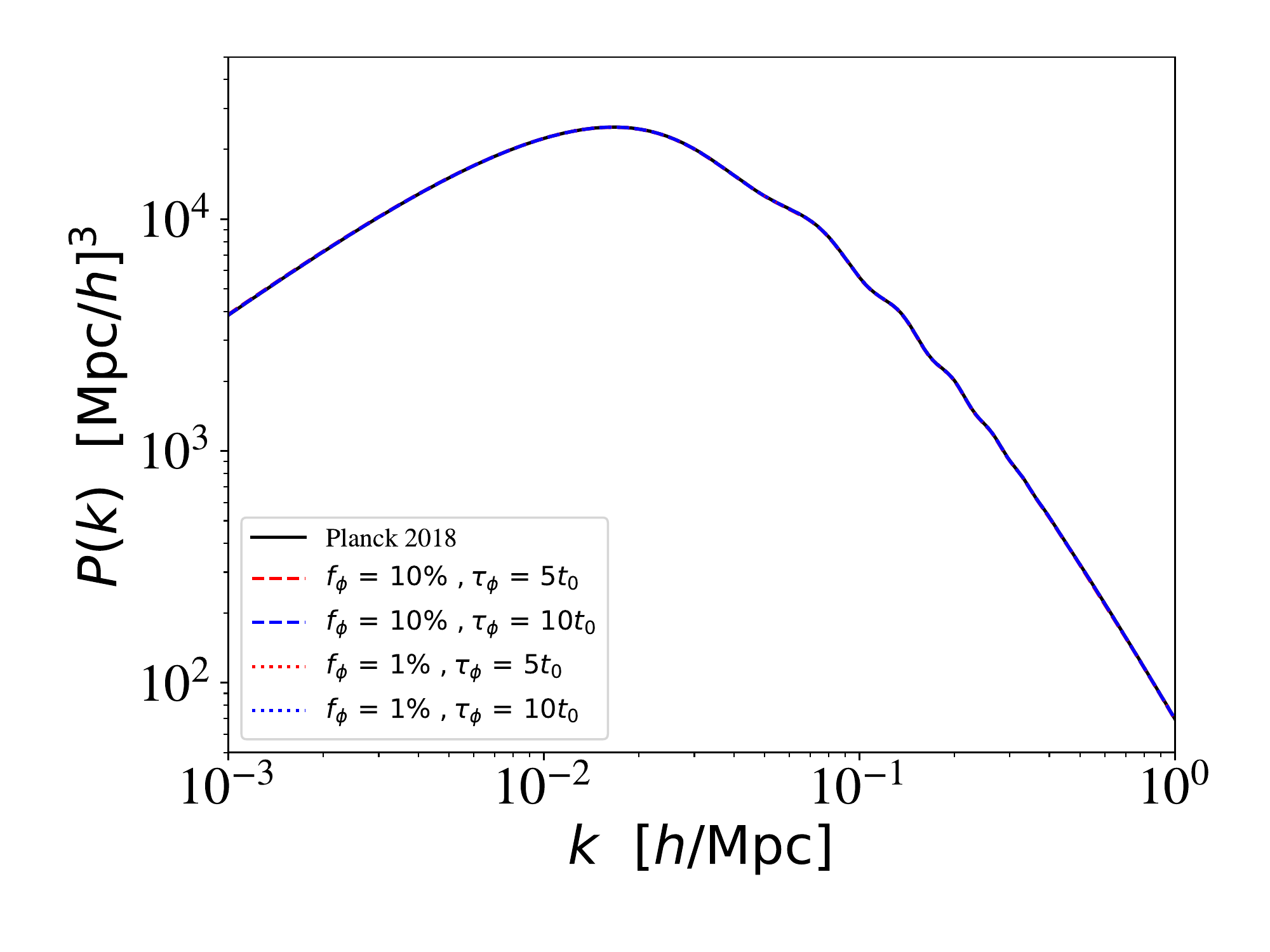}
\includegraphics[width=8cm,height=8cm]{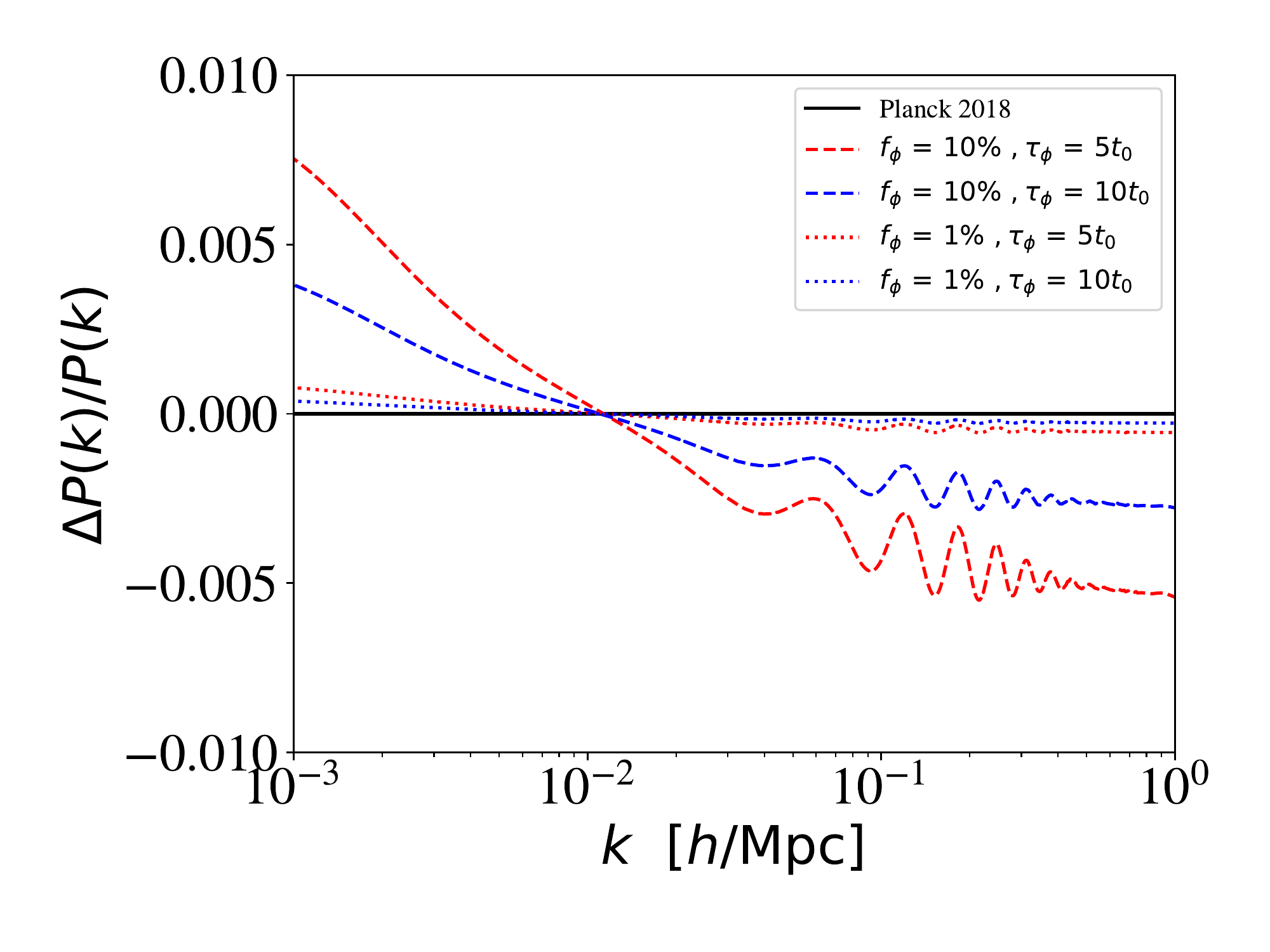}
\centering
\caption{Left: Matter power spectrum compared to Planck 2018 $\Lambda$CDM (black) for the four benchmark points. Right: magnitudes of the deviation over the $\Lambda$CDM value.}
\label{Pk}
\end{figure}

Fig.\ref{Cl} and Fig.\ref{Pk} show deviations in $C^{TT}_{\ell}$ and $P(k)$ from the $\Lambda$CDM baseline model (in black) respectively for four benchmark points 
extracted from Fig.\ref{rlight} which correspond to the explicit values $m_{\phi}=1.91$ keV, $\lambda=3.35\times 10^{-9}$ (red dotted); $m_{\phi}=1.35$ keV, $\lambda=3.98\times10^{-9}$ (blue dotted); $m_{\phi}=6.11$ keV, $\lambda=5.91\times 10^{-10}$ (red dashed); 
and $m_{\phi}=4.25$ keV, $\lambda=7.09\times 10^{-10}$ (blue dashed).
The first (later) two benchmark points have $f_{\phi}=10\%$  ($1\%$) and $\tau_{\phi}=\{5t_{0},10 t_{0}\}$ respectively. Explicitly, 
\begin{itemize}
\item For $C^{TT}_{\ell}$ a fixed $\theta_{s}$ defined as 
\begin{eqnarray}{\label{CMBa}}
\theta_{s}=\frac{r_{s}(z_{*})}{D_{A}(z_{*})},
 \end{eqnarray}
with $r_{s}(z)=\int^{\infty}_{z}c_{s}dz'/H(z')$ the sound horizon and $D_{A}(z)=\int^{z}_{0}dz'/H(z')$ the angular distance, 
implies that a smaller $\Omega_{\rm{cdm}}$ after recombination requires a larger $\Omega_{\Lambda}$,
where $z_{*}$ is the redshift at recombination.
This leads to an enhanced Integrated Sachs-Wolfe effect on $C^{TT}_{\ell}$ at large scales (small $\ell$ regions) similar to the case of DM decay to dark radiation \cite{Poulin:2016nat}.
As seen in Fig.\ref{Cl}, the deviations in $C^{TT}_{\ell}$ relative to $\Lambda$CDM are up to $\sim 12\%$ in the low $\ell$ region for the decaying DM component with $\tau_{\phi}\sim 5-10 t_{0}$ and $f_{\phi}\sim 10\%$.
These results are qualitatively consistent with \cite{Kumar:2018yhh} despite the fact that different best-fit values were used. 
Likewise, there is an enhanced Sachs-Wolfe effect on $C^{TT}_{\ell}$ at small scales (high $\ell$ regions) because of the additional radiation due to DM decay.
Finally, the DM decay gives rise to sub-dominant effects on CMB polarization \cite{Clark:2020miy}.
\item For $P(k)$ a larger $\Omega_{\Lambda}$ suppresses the growth factor $D(a)$ at small scales (large $k$ regions).
This is clearly seen in Fig.\ref{Pk}, 
where the deviations in $P(k)$ relative to $\Lambda$CDM are up to $\sim 0.5\%$ in the large $k$ regions  for $\tau_{\phi}\sim 5-10 t_{0}$ and  $f_{\phi}\sim 10\%$,
which is consistent with the results of \cite{Poulin:2016nat}.
The effect on $P(k)$ is instead enhanced at large scales (small $k$ regions), 
since $P(k)$ is proportional to $\Omega^{-2}_{\rm{cdm}}$ at scales $k\leq k_{\rm{eq}}$ with the subscript ``eq'' referring to the time of matter-radiation equality. 
The suppression on $P(k)$ in the large $k$ regions leads to a relatively mild suppression on $\sigma_{8}$ which is the value of $\sigma_{R}$ for $z=0$ and $R=8h^{-1}\rm{Mpc}$ with the definition \cite{WMAP:2003elm}
\begin{eqnarray}{\label{sigma8}}
\sigma_{R}(z)=\frac{1}{2\pi}\int^{\infty}_{0}k^{3}P(k,z)W^{2}_{R}(k)d \ln k,
 \end{eqnarray}
where $W_{R}(k)$ is the Fourier transform of window function.
\end{itemize}
As noted in \cite{Poulin:2016nat}, the enhancement on $C^{TT}_{\ell}$ in low $\ell$ regions and the suppression on $P(k)$ in large $k$ regions can be compensated by a larger $N_{\rm{eff}}$ simultaneously,
which implies certain degree of degeneracy between $N_{\rm{eff}}$ and the DM parameters.

\subsection{Hubble tension}
\begin{figure}
\centering
\includegraphics[width=16cm,height=16cm]{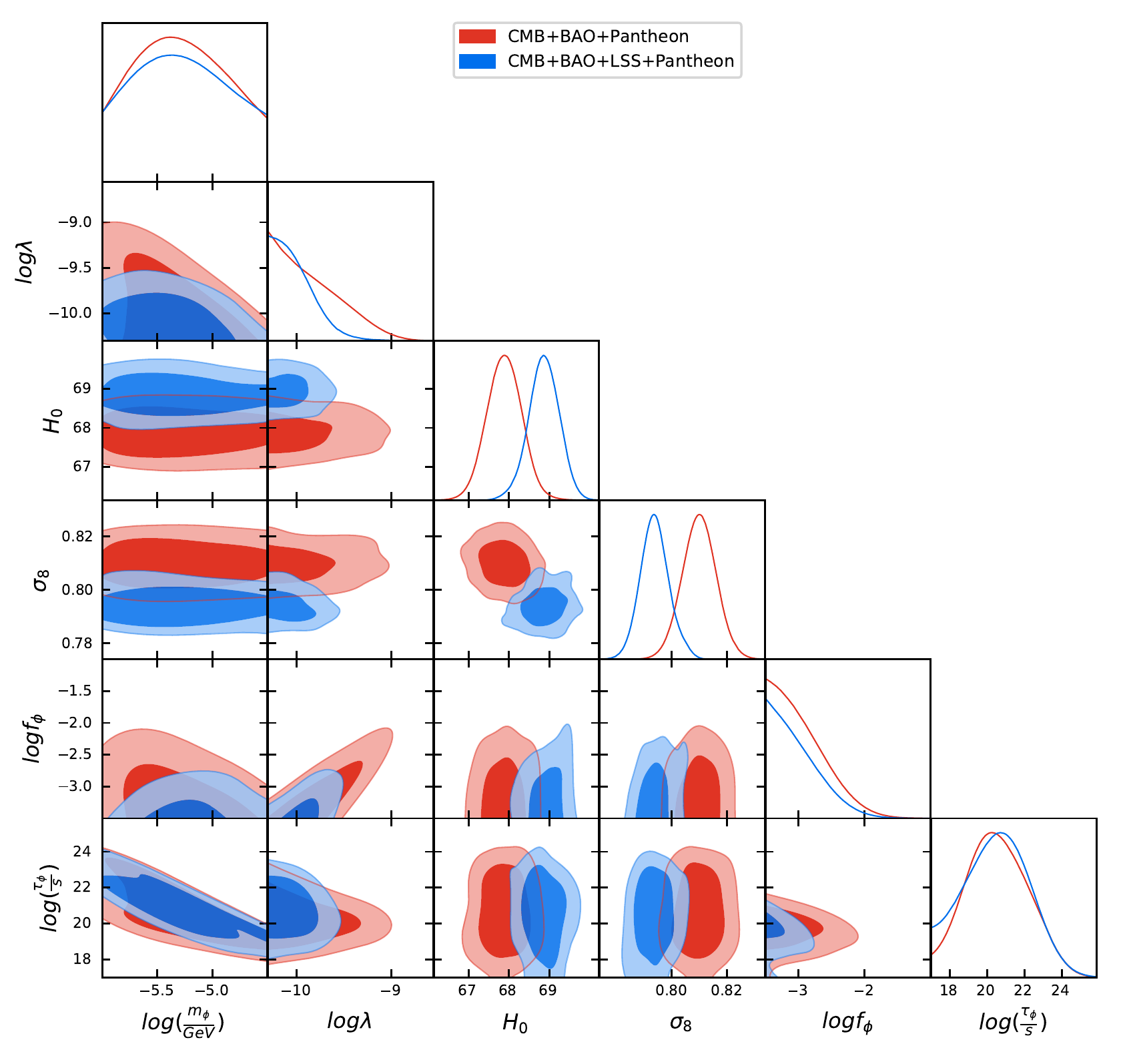}
\centering
\caption{The MCMC fit of our DM model within the hierarchical coupling scenario projected to a short chain of parameters composed of $\lambda$, $m_\phi$, $\tau_{\phi}$, $f_\phi$, $H_0$ and $\sigma_8$, 
with the Planck 2018+BAO(+LSS)+Pantheon data sets shown in red (blue). Parameter $H_0$ is in units of km s$^{-1}$Mpc$^{-1}$, $\tau_\phi$ in sec and $m_{\phi}$ in GeV.
For a complete result of the MCMC fit, see Fig.\ref{mcmc}.}
\label{mcmc2}
\end{figure}

Instead of fixing the cosmological parameters as above, 
we now make a MCMC fit of our DM model using MontePython \cite{Brinckmann:2018cvx}. 
Fig.\ref{mcmc2} shows the MCMC fit of the DM model within the hierarchical coupling scenario
projected to the short chain of parameters composed of $\lambda$, $m_\phi$, $\tau_{\phi}$, $f_\phi$, $H_0$ and $\sigma_8$, 
where the Planck 2018 TT+TE+EE+low$\ell$+lensing \cite{Planck:2018vyg}+BAO \cite{Beutler:2011hx, BOSS:2013rlg,Ross:2014qpa,BOSS:2013igd}+Pantheon \cite{Scolnic:2018} data sets and the Planck 2018 TT+TE+EE+low$\ell$+lensing from Planck 2018 \cite{Planck:2018vyg}+BAO \cite{Beutler:2011hx, BOSS:2013rlg,Ross:2014qpa,BOSS:2013igd}+LSS \cite{Heymans:2013fya, Planck:2013lkt}+Pantheon \cite{Scolnic:2018} data sets are shown in red and blue, respectively.  
We present a more complete result of the MCMC fit in Fig.\ref{mcmc} in the Appendix.

Table \ref{bvalue} shows the best-fit values of the cosmological parameters, where one finds
\begin{eqnarray}{\label{bfh}}
\rm{CMB+BAO(+LSS)+Pantheon}: H_{0}=68.31(69.34)~\rm{km~s}^{-1}\rm{Mpc}^{-1}.
 \end{eqnarray}
Compared to the value of $H_{0}=(73.20\pm 1.30)$ km s$^{-1}$Mpc$^{-1}$ at 68$\%$ CL reported by the local experiments \cite{Riess:2020fzl}, 
the significance of Hubble tension is now of order $\sim 3.8(3.0)\sigma$.
Without imposing the BAO data set,
the ability of relaxing the $H_0$ tension can be more obviously enhanced.
Because BAO data strongly constrains the value of $H_{0}r_{s}$, 
which implies that an increase in $r_{s}$ due to the late-time DM decay requires a reduction in $H_0$.
In Fig.\ref{mcmc2} the resolution of Hubble tension favors the parameter regions of $\lambda\sim 10^{-10}$ and $m_{\phi} \sim 1-10$ keV, 
which point to a percent level of $f_{\phi}$ and $\tau_{\phi}\sim 10~t_{0}$.

\begin{table}
\begin{center}
\begin{tabular}{ccc}
\hline\hline
$\rm{Parameters}$~ & Planck 2018+BAO+Pantheon  ~& Planck 2018+BAO+LSS+Pantheon \\ \hline
$10^{2}\omega_{b}$ & $2.26273$ &  $2.26884$ \\
$\omega_{\rm{cdm}}^{\rm{ini}}$ & $0.118102$ & $0.115725$ \\
$100\theta_{s}$ & $1.04192$ & $1.04226$\\
$\ln{(10^{10}A_{s})}$ & $3.04992$ &  $3.02874$\\
$n_{s}$ & $0.971473$  & $0.976201$ \\
$\tau_{\rm{reio}}$ & $0.057636$  & $0.0520685$ \\ 
$H_{0}$ & $68.3097$ & $69.3441$ \\
$\sigma_8$ & $0.807764$ & $0.792412$  \\
$\Omega_{m}$ & $0.302971$ & $ 0.289185$ \\
$r_{\rm{drag}}$ & $147.308$  & $147.877$ \\
\hline \hline
\end{tabular}
\caption{The best-fit values of the cosmological parameters in our DM model within the hierarchical coupling scenario with respect to the $\rm{CMB+BAO}$(+\rm{LSS})+Pantheon data sets,
which lead to $\Delta\chi^{2}=-1.3$ (-1.5) relative to the $\Lambda$CDM model by following the $\chi^2$ criteria in \cite{Schoneberg:2021qvd}.}
\label{bvalue}
\end{center}
\end{table}

Apart from the $H_0$ tension, the MCMC fit to Planck 2018 TT+TE+EE+low$\ell$+lensing + BAO+ LSS+Pantheon  data sets in Table \ref{bvalue} is able to give a smaller best-fit value 
 \begin{eqnarray}{\label{sigma8}}
\sigma_{8}=0.792,
 \end{eqnarray}
compared to $\sigma_{8}=0.812$ \cite{Planck:2018vyg} in the Planck 2018 $\Lambda$CDM.
Referring to the value of $\sigma_{8}=0.75\pm 0.03$ at 68$\%$ CL reported by the LSS \cite{Planck:2013lkt}, 
this result verifies that the $\sigma_8$ tension between the Planck and LSS data can be mildly reduced in our model as a byproduct.

\section{Stellar cooling constraints}
\label{astro}
The direct couplings of $\phi$ to the SM charged leptons with $m_{\phi}$ of order keV in our model enable $\phi$ to be produced in stellar systems \cite{Raffelt1996} such as the Sun, red giants (RGs) and white dwarfs (WDs).\footnote{We neglect the relatively weaker SN1987A limits.} 
Each of these astrophysical objects provides a local thermal bath with a characteristic temperature.
Since $m_{\phi}$ is comparable with the characteristic temperatures of these stellar systems, 
a large number of $\phi$ particles are produced without a Boltzmann suppression, 
which contributes to a new form of stellar energy loss after they escape the core of the stellar system as a result of the feeble interactions with the thermal bath therein. 
Such new stellar cooling allows us to place constraints on the feeble interactions far stronger than in ground-based experiments.

\subsection{Universal coupling scenario}
Let us firstly consider the main production for $\phi$ in the universal coupling scenario.
In this situation the coupling of $\phi$ to electrons plays a key role,
which suggests that 
\begin{itemize}
\item the electron-positron inverse decay $e\bar{e}\rightarrow \phi$
\item the Bremsstrahlung emission of $\phi$ in $e-X$ scattering with $X$ being nucleus,
\item the Compton-like scattering process $e\gamma\rightarrow e\phi$,
\end{itemize}
are important.  
The luminosity per volume in the Bremsstrahlung emission calculated by Ref.\cite{Hardy:2016kme}
implies that $\lambda_{e} \leq 7\times 10^{-16}$ for RGs, which has already excluded the parameter space preferred by the Hubble tension within this coupling scenario.

\begin{figure}
\centering
\includegraphics[width=16cm,height=6cm]{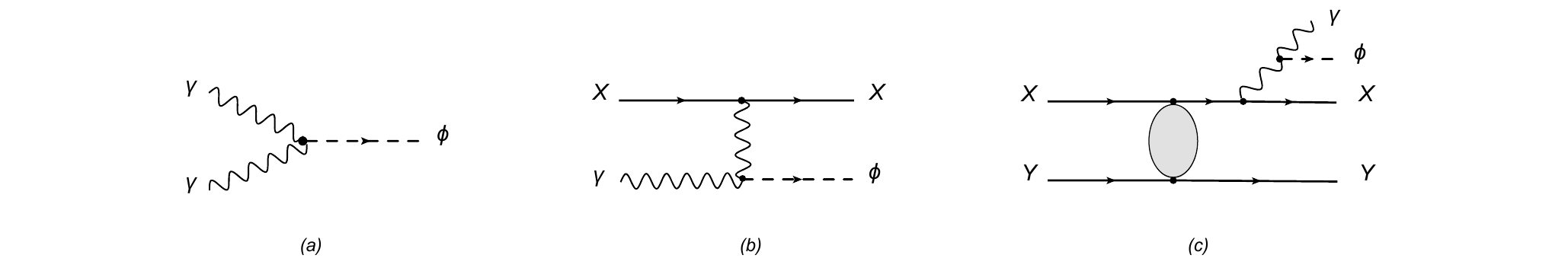}
\centering
\caption{Feynman diagrams for the $\phi$ production in stellar systems: $(a)$ photon inverse decay, $(b)$ Primakoff emission, and $(c)$ Bremsstrahlung emission.}
\label{Feyn}
\end{figure}

\subsection{Hierarchical coupling scenario}
Unlike in the universal coupling scenario where the Yukawa coupling $\lambda_{e}$ is the critical parameter,  
the dominant production for $\phi$ in the hierarchical coupling scenario is instead determined by the Yukawa coupling $\lambda_{\tau}$. 
It gives rise to an effective coupling of $\phi$ to di-photons,
implying that 
\begin{itemize}
\item the photon inverse decay $\gamma\gamma \rightarrow \phi$,
\item the Primakoff process $X\gamma \rightarrow X\phi$ with $X$ being nucleus or electron,
\item the Bremsstrahlung emission of $\phi$ in $X+Y\rightarrow X+Y+\phi+\gamma$ with $X$ and $Y$ being nucleus
\end{itemize}
are the main processes. The Feynman diagrams with respect to these processes are shown in Fig.\ref{Feyn}.

$(a)$. In the photon inverse decay process, we follow the treatment on majoron ($J$) production via the neutrino inverse decay process $\bar{\nu}\nu\rightarrow J$ \cite{Choi:1989hi, Kachelriess:2000qc, Farzan:2002wx,Heurtier:2016otg}. 
The luminosity per volume is given by 
\begin{eqnarray}{\label{ann}}
\frac{dL_{a}}{dV}=\int d\Pi_{p_{1}}d\Pi_{p_{2}}d\Pi_{p_{\phi}}\left[\mid\mathcal{M}_{a}\mid^{2}E_{\phi}f(E_{1})f(E_{2})\right](2\pi)^{4}\delta^{(4)}(p_{1}+p_{2}-p_{\phi})
\end{eqnarray}
with 
\begin{eqnarray}{\label{Ma}}
 \mid\mathcal{M}_{a}\mid^{2}=\frac{2\alpha^{2}\lambda^{2}}{9\pi^{2}m^{2}_{\tau}} (p_{1}\cdot p_{2})^{2},
\end{eqnarray}
where $p_{1}=(E_{1},\mathbf{p}_{1})$ and $p_{2}=(E_{2},\mathbf{p}_{2})$ are the momenta of the two incoming photons, $p_{\phi}=(E_{\phi}, \mathbf{p}_{\phi})$ is the momentum of the outgoing $\phi$, $\mathcal{M}_a$ is the annihilation amplitude, and $f(E)=(e^{E/T}-1)^{-1}$ is the Bose-Einstein distribution function.

$(b)$. In the the Primakoff process, we refer to a new scalar production via its coupling to photons \cite{Balaji:2022noj}. 
The luminosity per volume is
\begin{eqnarray}{\label{pri}}
\frac{dL_{p}}{dV}=\sum_{X}2n_{X}\int \frac{d^{3}\mathbf{k}_{\gamma}}{(2\pi)^{3}}\sigma_{p}E_{\phi}f(E_{\gamma})
\end{eqnarray}
with
\begin{eqnarray}{\label{Mp}}
\sigma_{p}=64\pi\alpha Z^{2}_{X}\frac{E_{\gamma}\Gamma_{\phi}}{m^{2}_{\phi}}\frac{\sqrt{E^{2}_{1}-m^{2}_{\phi}}(E_{\gamma}-m_{\phi})}{(m^{2}_{\phi}+2m_{\phi}E_{\gamma}+k^{2}_{s})^{2}},
\end{eqnarray}
where $Z_X$ is the atomic number of the nucleus, 
$k_{\gamma}=(E_{\gamma},\mathbf{k}_{\gamma})$  is the momentum of the incoming photon, $k_s$ is the screening scale \cite{Balaji:2022noj}, 
$\sigma_p$ is the scattering cross section, and $f$ is the Bose-Einstein distribution function.

\begin{table}
\begin{center}
\begin{tabular}{cccccc}
\hline\hline
$\rm{Star}$~ &\rm{Core~component} ~ & $T_c$ [keV] ~& $n_{e}$ [cm$^{-3}$] ~&~$R$ [cm]~& $\mathcal{L}/\mathcal{L}_{\odot}$ \\ \hline
Sun & $75\%~$H, $25\%~^{4}$He & 1 & $10^{26}$ & $7\times10^{10}$ & $0.03$\\
RGs & $^{4}$He & 10 & $3\times10^{27}$ & $3\times10^{9}$ & $2.8$\\
WDs & $50\%~^{12}$C, $50\%~^{16}$O & 6 & $10^{30}$ & $10^{9}$ & $0.03$\\
\hline \hline
\end{tabular}
\caption{Stellar parameters \cite{Balaji:2022noj} for the Sun, RGs and WDs, with $T_{c}$ the stellar core temperature, $n_e$ the number density of electron, $R$ the radius, and $\mathcal{L}$ the stellar cooling limit.}
\label{stp}
\end{center}
\end{table}

$(c)$. The Bremsstrahlung emission of $\phi$ mimics the DM production through dark photon \cite{Chang:2018rso}. The luminosity per volume reads as 
\begin{eqnarray}{\label{bre}}
\frac{dL_{b}}{dV}&=&\int d\Pi_{p_{1}}d\Pi_{p_{2}}d\Pi_{p_{3}}d\Pi_{p_{4}}d\Pi_{k_{\gamma}}d\Pi_{p_{\phi}} (2\pi)^{4}\delta^{(4)}(p_{1}+p_{2}-p_{3}-p_{4}-p_{\phi}-k_{\gamma}) \nonumber\\
&\times&\left[\mid\mathcal{M}_{b}\mid^{2}E_{\phi}f(E_{1})f(E_{2})\right]
\end{eqnarray}
with 
\begin{eqnarray}{\label{Mb}}
\mid\mathcal{M}_{b}\mid^{2}&=&\frac{4\alpha^{3}\lambda^{2}}{9\pi m^{2}_{\tau}}\mid\mathcal{M}\mid^{2}_{np}
\times\left(\frac{2p^{\mu}_{1}}{k^{2}-2p_{1}\cdot k}+\frac{2p^{\mu}_{3}}{k^{2}+2p_{3}\cdot k}\right)\left(\frac{2p^{\alpha}_{1}}{k^{2}-2p_{1}\cdot k}+\frac{2p^{\alpha}_{3}}{k^{2}+2p_{3}\cdot k}\right)\nonumber\\
&\times& \frac{g_{\mu\nu}g_{\alpha\beta}g_{\rho\delta}}{k^{4}}
\times\left(k^{\nu}k^{\rho}_{\gamma}-g^{\nu\rho}k\cdot k_{\gamma}\right) 
\left(k^{\beta}k^{\delta}_{\gamma}-g^{\beta\delta}k\cdot k_{\gamma}\right)
\end{eqnarray}
where $\mid\mathcal{M}\mid^{2}_{np}$ \cite{Rrapaj:2015wgs} is the squared amplitude for the process with no bremsstrahlung,
$p_{1}$ and $p_2$ are the momenta of the incoming nucleus with energy $E_1$ and $E_2$ respectively, 
$p_{3}$ and $p_4$ are the momenta of the outgoing nucleus, 
while $k_{\gamma}$, $k$ and $p_{\phi}=(E_{\phi}, \mathbf{p}_{\phi})$ are the momenta of the outgoing photon, off-shell photon and $\phi$ respectively.
Here, $f(E)$ is the Fermi-Dirac distribution function.
In this process it is valid to treat the nucleus as non-relativistic particles. 

\begin{figure}
\centering
\includegraphics[width=15cm,height=11cm]{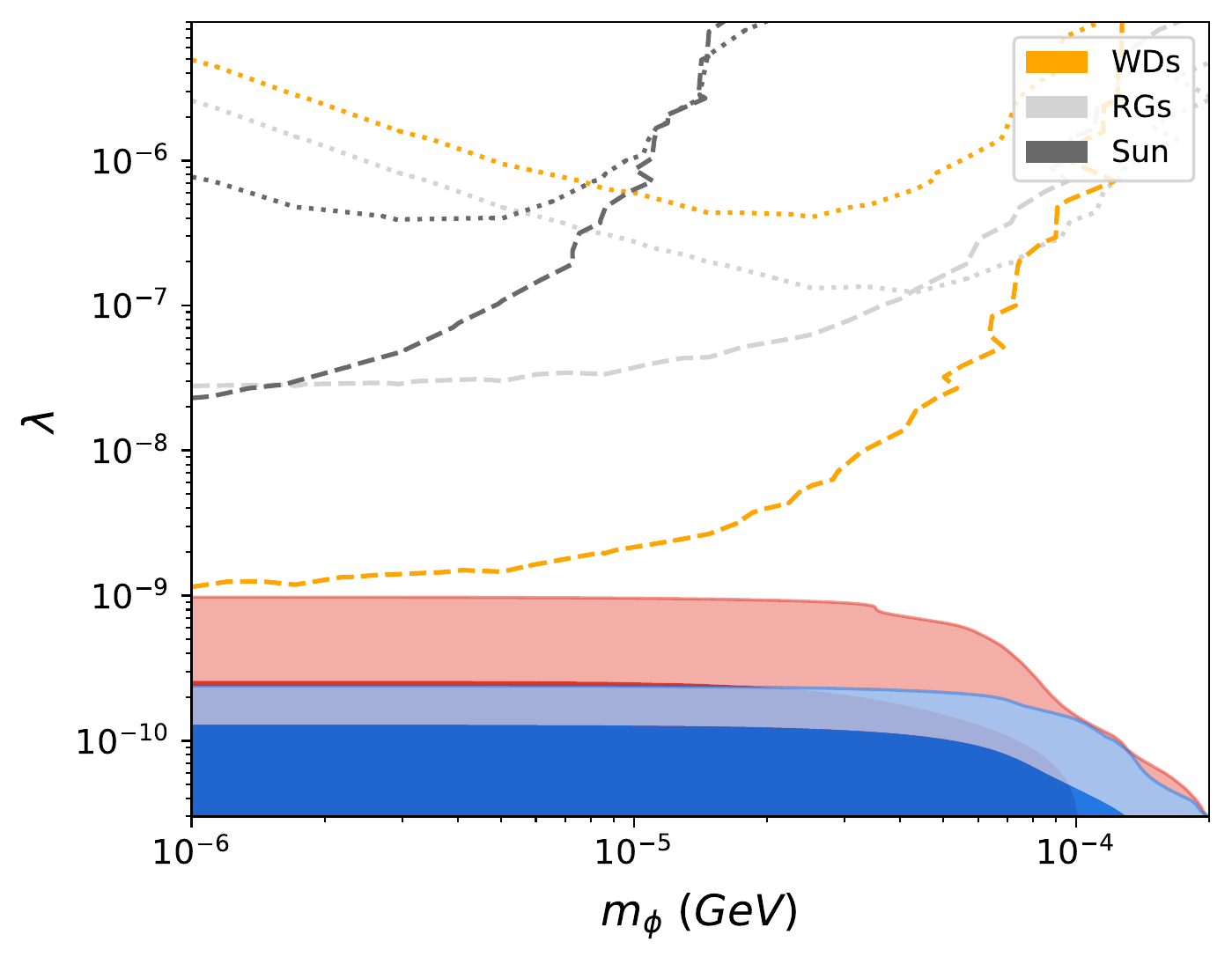}
\centering
\caption{Stellar cooling bounds on the parameter space with respect to reducing the Hubble tension in Fig.\ref{mcmc2}, 
through $(a)$ photon inverse decay (dot) and $(b)$ Primakoff emission (dashed) of $\phi$ in the Sun (black), RGs (gray) and WDs (brown) respectively. Regions above the bounds are excluded. See text for details.}
\label{ps}
\end{figure}

Comparing the new emission rates in Eq.(\ref{ann}), Eq.(\ref{pri}) and Eq.(\ref{bre}) to the stellar cooling data of the Sun, RGs, and WDs as shown in Table.\ref{stp}, we place constraints on the parameter space with respect to reducing the Hubble tension in Fig.\ref{ps}, where we adopt the criteria that the new luminosity is less than $\sim 0.03\mathcal{L}_{\odot}$, $\sim 2.8\mathcal{L}_{\odot}$  and $\sim 0.03\mathcal{L}_{\odot}$ in the Sun, RGs, WDs respectively, with $\mathcal{L}_{\odot}$ the luminosity of Sun.
Here, we explicitly show the limits from $(a)$ photon inverse decay in dot and $(b)$ Primakoff emission in dashed whereas the Sun, RGs and WDs are referred to as black, gray and brown respectively.
Given a stellar the process $(c)$ gives much weaker constraints compared to the processes $(a)$ and $(b)$, which has been neglected in Fig.\ref{ps}.
It turns out the Primakoff emission of $\phi$ in the WDs offers the most stringent limit.
Unlike the Sun whose luminosity is fixed, the luminosities of WDs have a large uncertainty which spans a few orders of magnitude. 
If one adjusts $\mathcal{L}$ to be lower than the reference value in Table.\ref{stp} by two orders, 
the upper bound on $\lambda$ in Fig.\ref{ps} will be lowered by one order, 
suggesting that the parameter space is nearly excluded.
In this sense, our model can be tested by the future observations of WDs made by SDSS \cite{Ourique:2018,Kepler:2019} and Gaia \cite{Bergeron:2019,Fusillo:2021}.

The derived stellar cooling limits in Fig.\ref{ps} are subject to uncertainties due to a few simplifications - neither the polarization effect of photon in the medium nor the dependence of the stellar parameters on the radius has been taken into account.

\section{Conclusion}
\label{con}
In this paper we have studied a new type of freeze-in DM model through the SM lepton portal. 
Given the feeble interactions, such DM is produced from the SM thermal bath via the freeze-in mechanism,
then decays to photons in the late-time Universe with its lifetime of order the age of Universe.
Therefore, our model serves as a realistic realization of DM with late-time decay which alleviates the Hubble tension.
Based on the MCMC analysis on this model with the hierarchical coupling scenario,
we have shown the best-fit value of $H_{0}=68.31(69.34)$ km s$^{-1}$Mpc$^{-1}$ with respect to Planck 2018+BAO(+LSS)+Pantheon data sets in the parameter regions referring to $f_{\phi}\sim 1\%$ and $\tau_{\phi}\sim 10~t_{0}$, 
which suggests the significance of Hubble tension of order $\sim 3.8 (3.0)\sigma$.

We have also used the complimentary stellar cooling data to set stringent constraints on the parameter space with respect to the Hubble tension. The analysis is highly unlikely unless one specifies the DM model as we do. 
We have shown that while the universal coupling scenario has been excluded, 
according to our quantitative analysis on the photon inverse decay, Primakoff emission, and Bremsstrahlung emission of $\phi$ in the representative stellar systems 
the hierarchical coupling scenario can be tested by the future observations of WDs made by SDSS and Gaia.

Finally, we emphasize two points left for future work.
First, one can use the Lyman-$\alpha$ data \cite{Garzilli:2019qki,Villasenor:2022aiy,Capozzi:2023xie} to constrain this DM model, after taking into account the fact that $\phi$ is only a subdominant fraction of the observed DM.
The second point is that besides the future observations of WDs X-ray telescopes may be also used to test this DM component.
Having said so $\phi$ cannot be used to address the 3.5 keV X-ray reported by \cite{Bulbul:2014sua,Boyarsky:2014jta}.

\section*{Acknowledgements}
This research is supported in part by the National Natural Science Foundation of China under Grant No. 11775039,
the High-level Talents Research and Startup Foundation Projects for Doctors of Zhoukou Normal University (ZKNUC2021006) and Scientific research projects of universities in Henan Province, China (23A140027).

\appendix
\section{The complete fit}
\begin{figure}
\centering
\includegraphics[width=16.5cm,height=19cm]{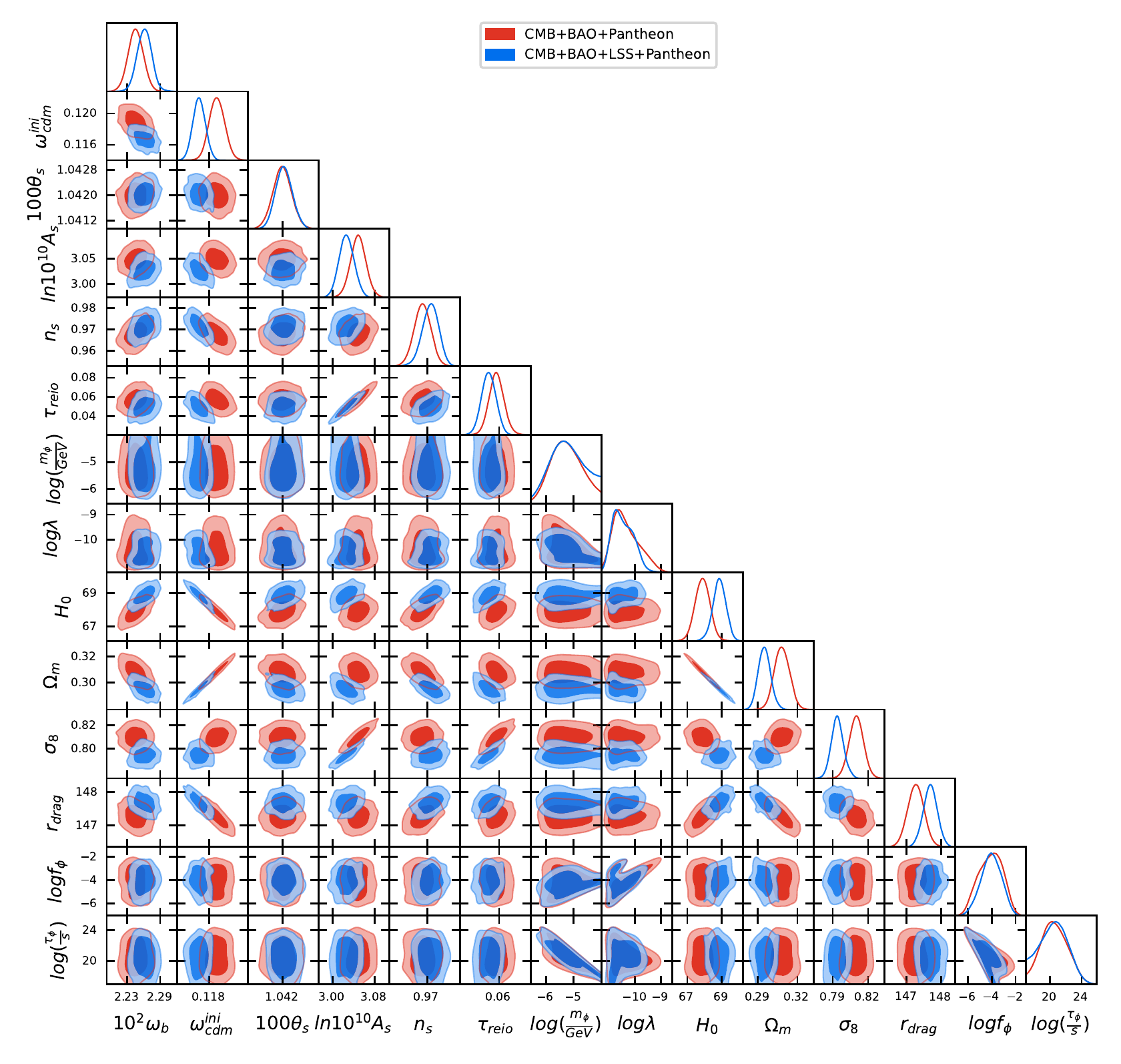}
\centering
\caption{MCMC fit of the DM model within the hierarchical coupling scenario to the Planck 2018+BAO(+LSS)+Pantheon data sets as shown in red (blue).
We have taken $N_{\rm{eff}}=3.044$, and $r_{\rm{drag}}$ is in units of Mpc.}
\label{mcmc}
\end{figure}

Compared to Fig.\ref{mcmc2} where the values of the DM model parameters, $H_0$ and $\sigma_8$ are highlighted, 
Fig.\ref{mcmc} shows the complete result of the MCMC fit to our DM model with respect to the Planck 2018+ BAO(+LSS)+Pantheon data sets.

\end{document}